\documentclass[Physsubmission, Phys]{SciPost}

\hypersetup{
    colorlinks,
    linkcolor={red!50!black},
    citecolor={blue!50!black},
    urlcolor={blue!80!black}
}

\usepackage{graphicx}
\usepackage[bitstream-charter]{mathdesign}
\DeclareSymbolFont{usualmathcal}{OMS}{cmsy}{m}{n}
\DeclareSymbolFontAlphabet{\mathcal}{usualmathcal}

\begin{document}

\begin{center}{\Large \textbf{
Heavy Flavor and Jet Studies for the Future Electron-Ion Collider to Explore the Hadronization Process \\
}}\end{center}

\begin{center}
Xuan Li\textsuperscript{1$\star$}
\end{center}

\begin{center}
{\bf 1} Physics Division, Los Alamos National Laboratory
\\
* xuanli@lanl.gov
\end{center}

\begin{center}
\today
\end{center}


\definecolor{palegray}{gray}{0.95}
\begin{center}
\colorbox{palegray}{
  \begin{tabular}{rr}
  \begin{minipage}{0.1\textwidth}
    \includegraphics[width=22mm]{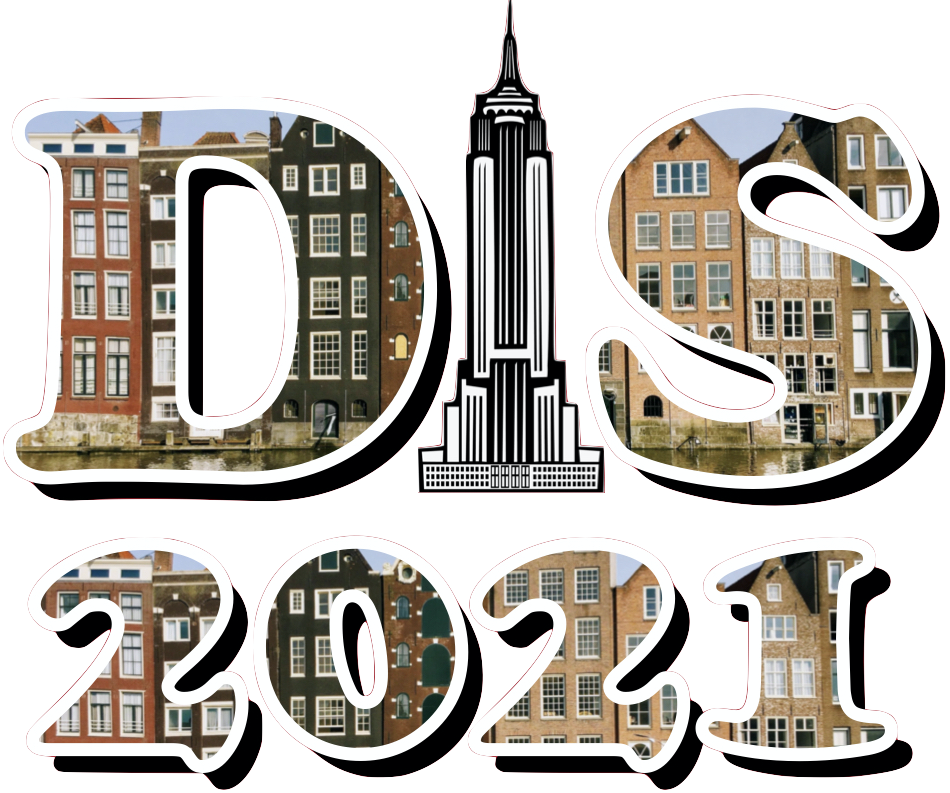}
  \end{minipage}
  &
  \begin{minipage}{0.75\textwidth}
    \begin{center}
    {\it Proceedings for the XXVIII International Workshop\\ on Deep-Inelastic Scattering and
Related Subjects,}\\
    {\it Stony Brook University, New York, USA, 12-16 April 2021} \\
    \doi{10.21468/SciPostPhysProc.}\\
    \end{center}
  \end{minipage}
\end{tabular}
}
\end{center}

\section*{Abstract}
{\bf
Heavy flavor production at the future Electron-Ion Collider (EIC) will allow us to precisely determine the quark/gluon fragmentation processes in vacuum and the nuclear medium especially within the poorly constrained kinematic region. Heavy flavor hadron and jet reconstruction with the recent EIC detector design have been studies in simulation. Results of corresponding physics projections such as the flavor dependent hadron nuclear modification factor $R_{eA}$ in electron+nucleus collisions will be shown. The statistical precision obtained by these proposed heavy flavor measurements for the future EIC provides a strong discriminating power in separating different theoretical predictions.
}


\section{Introduction}
\label{sec:intro}
One of the EIC science drivers is to explore how quarks and gluons form colorless hadrons, which is known as the hadronizaiton process. Such process can not be directly calculated by the perturbative Quantum Chromodynamics (pQCD) theory, and it relies on the extrapolation of global fits on experimental measurements. The future EIC will operate high luminosity electron+proton and electron+nucleus collisions with a variety of different nuclear species (mass number from 2 to 208) at center of mass energies from 20 to 140 GeV. A cleaner environment can be provided by the EIC compared to heavy ion collisions for exploring the hadronization process within both vacuum and a nuclear medium. Heavy flavor quarks (i.e. charm and bottom quarks) have different production mechanisms from light flavor quarks and are expected to experience different hadronization processes due to their mass differences ($m_{c,b} > \Lambda_{c} > m_{u,d,s}$) \cite{HF_had_th}. Heavy flavor nuclear modification factor $R_{eA}$ can help extracting the information about the hadronization in medium especially when the final state hadron carries a large momentum fraction relative to the parent parton. Simulation studies of reconstructed D-mesons and B-mesons with the help of a proposed forward silicon vertex/tracking detector together with the associated physics projections at the future EIC \cite{EIC_YR} will be discussed.

\begin{figure}[htb!]
\centering
\includegraphics[width=0.38\textwidth]{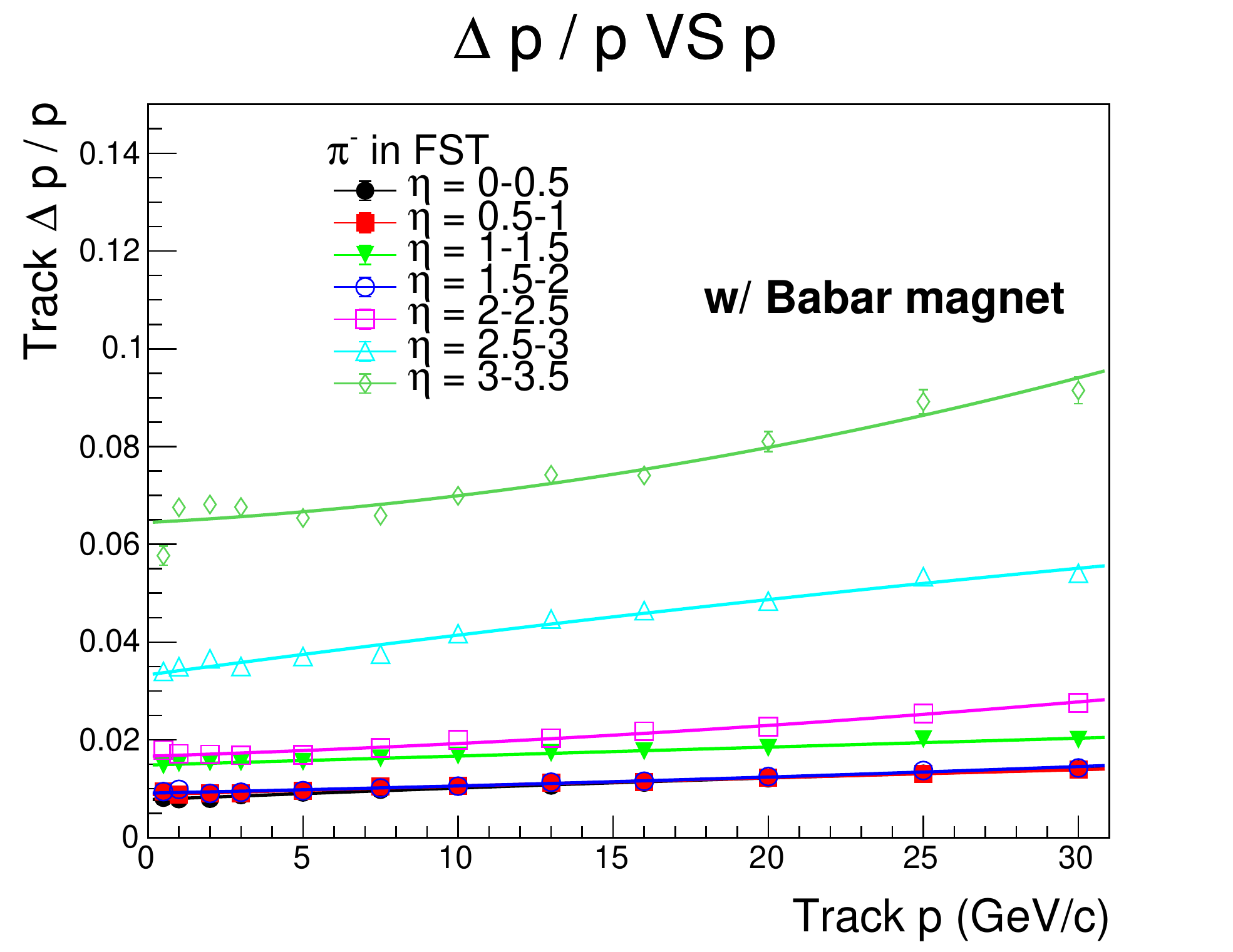}
\includegraphics[width=0.38\textwidth]{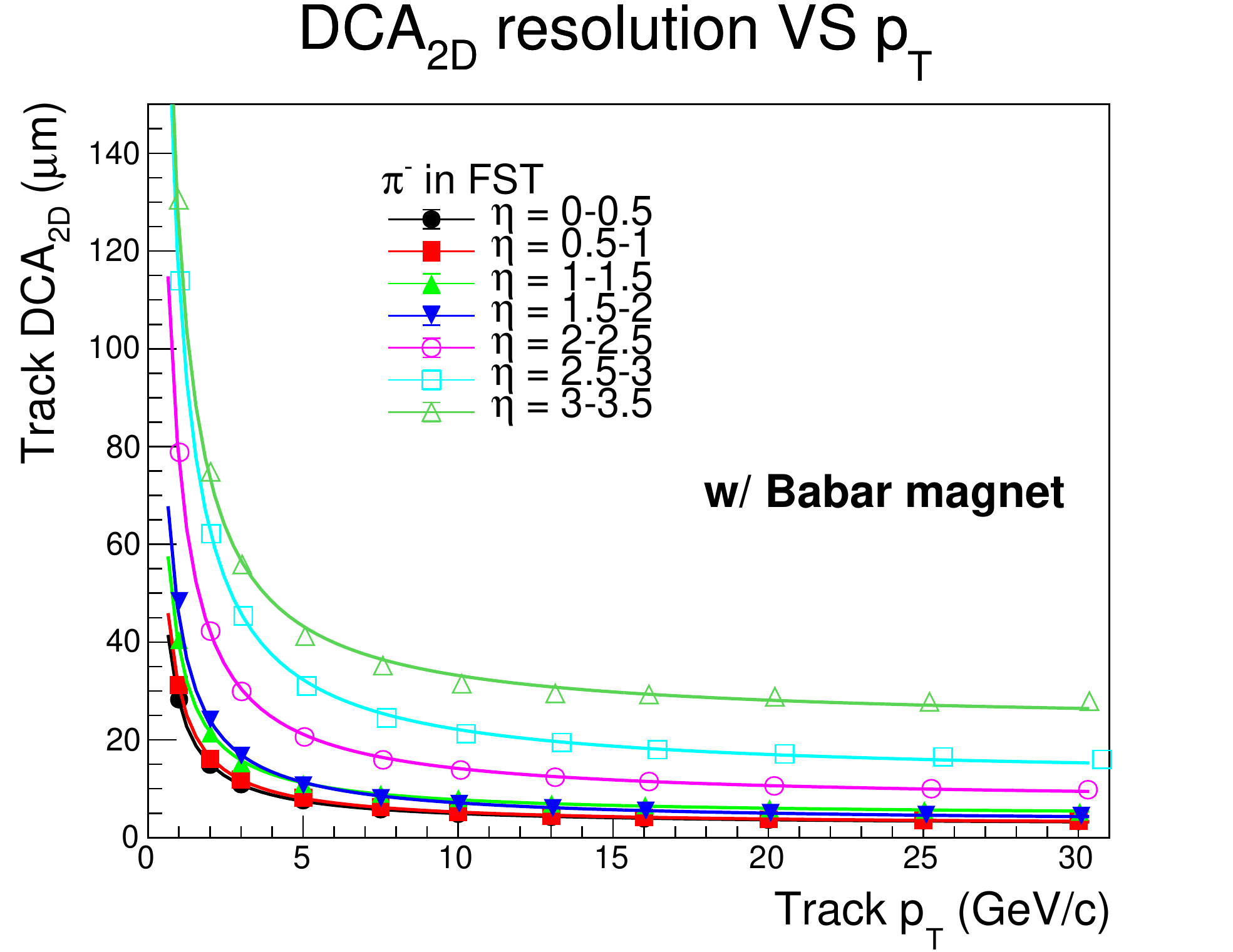}
\includegraphics[width=0.38\textwidth]{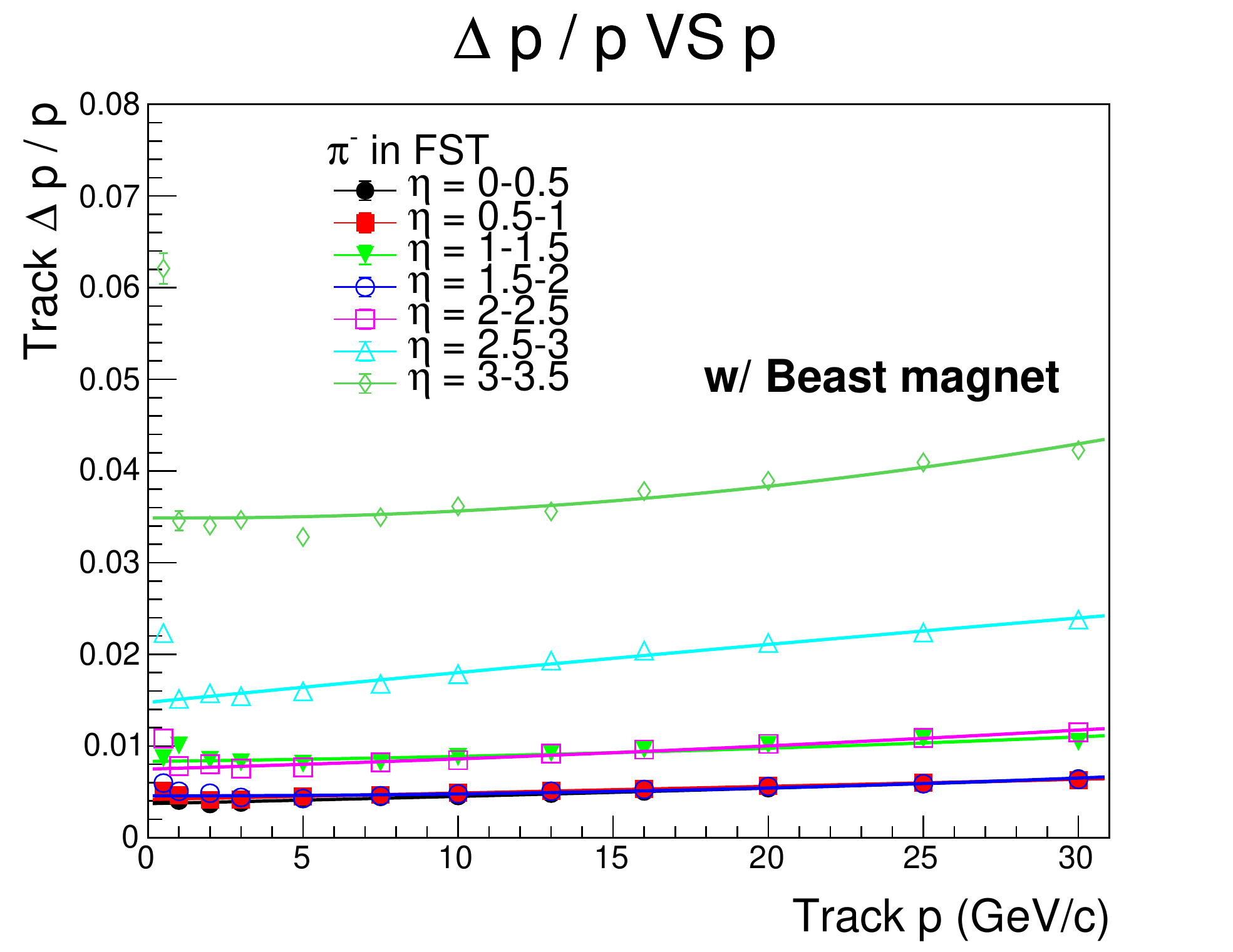}
\includegraphics[width=0.38\textwidth]{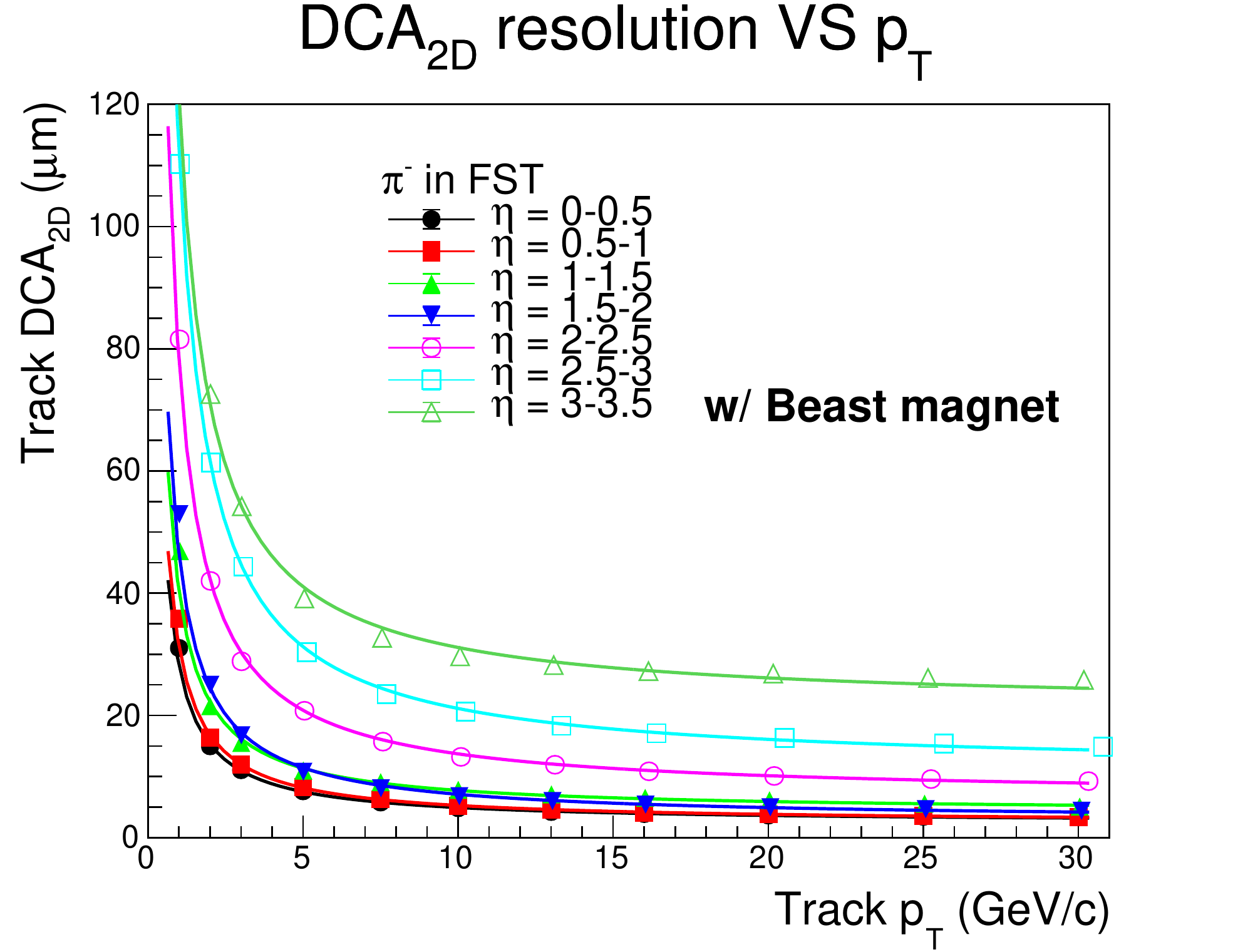}
\caption{Performance of an integrated silicon tracking system with the Babar magnet for the future EIC \cite{EIC_YR} is shown in the top and the results with the Beast magnet is shown in the bottom. The track momentum resolution versus track momentum in different pseudorapidities is shown in the left and the pseudorapidity dependent track transverse distance of closet approach resolution ($DCA_{2D}$) versus transverse momentum is illustrated in the right.}
\label{fig:detector_perfom}
\end{figure}

\section{Heavy Flavor Hadron and Jet Reconstruction in e+p Simulation}
\label{sec:reco}
To identify particles from heavy flavor decay in the hadron beam going direction at the EIC, a proposed Forward Silicon Tracker (FST), which can provide precise track reconstruction and Distance of Closest Approach (DCA) determination in the $1.0<\eta<3.5$ region, is under design and detector R$\&$D \cite{lanl_fst}. The performance of the proposed FST and other EIC detector sub-systems \cite{EIC_YR, Arrington:2021yeb} have been included in the simulation framework, which contains the event generation in PYTHIA8 \cite{Sjostrand:2007gs}, beam remnant background and the hadron/jet reconstruction chain. The tracking performance with two magnet options: Babar and Beast have been studied in GEANT4 \cite{AGOSTINELLI2003250} within the Fun4All framework. Figure~\ref{fig:detector_perfom} shows the momentum dependent tracking momentum resolution and transverse momentum dependent transverse DCA ($\rm{DCA}_{2D}$) resolution in different pseudorapidity regions. 

According to the factorization mechanism, cross section of the hadron with a flavor $b$ in $e+p$ collision can be described as Equation \ref{eq1}.
\begin{equation}
\frac{d\sigma_{e+p}^{b}}{dp_{T}d\eta} = \sum_{a}f_{a}(x_{BJ}, Q^{2})\cdot \sigma_{\gamma^{*} a \rightarrow b} \cdot D_{b}(z_{h}, \nu)  , 
\label{eq1}
\end{equation}
where $f_{a}(x_{BJ}, Q^{2})$ is the parton distribution function of a parton with flavor $a$ and carries the longitudinal momentum fraction $x_{BJ}$ relative to the parent proton, $\sigma_{\gamma^{*} a \rightarrow b}$ is the deeply inelastic scattering partonic process which can be precisely calculated by pQCD and $D_{b}(z_{h}, \nu)$ is the fragmentation function of a parton with flavor b that produces a final hadron carrying the momentum fraction $z_{h}$ relative to the parton. Reconstructed hadrons with different flavors/masses or within different pseudorapidities in $e+p$ collisions are sensitive to probed parton distribution functions in different Bjorken-x ($x_{BJ}$) regions. The accessed fragmentation functions also contain different hadron momentum fraction $z_{h}$ values.

\begin{figure}[htb!]
\centering
\includegraphics[width=0.9\textwidth]{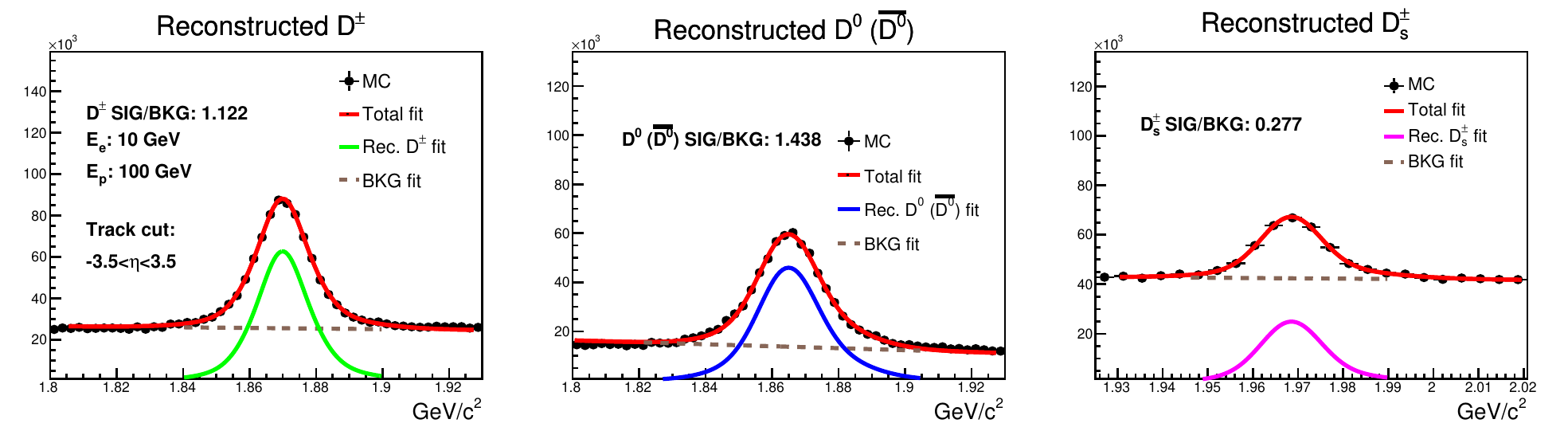}
\includegraphics[width=0.3\textwidth]{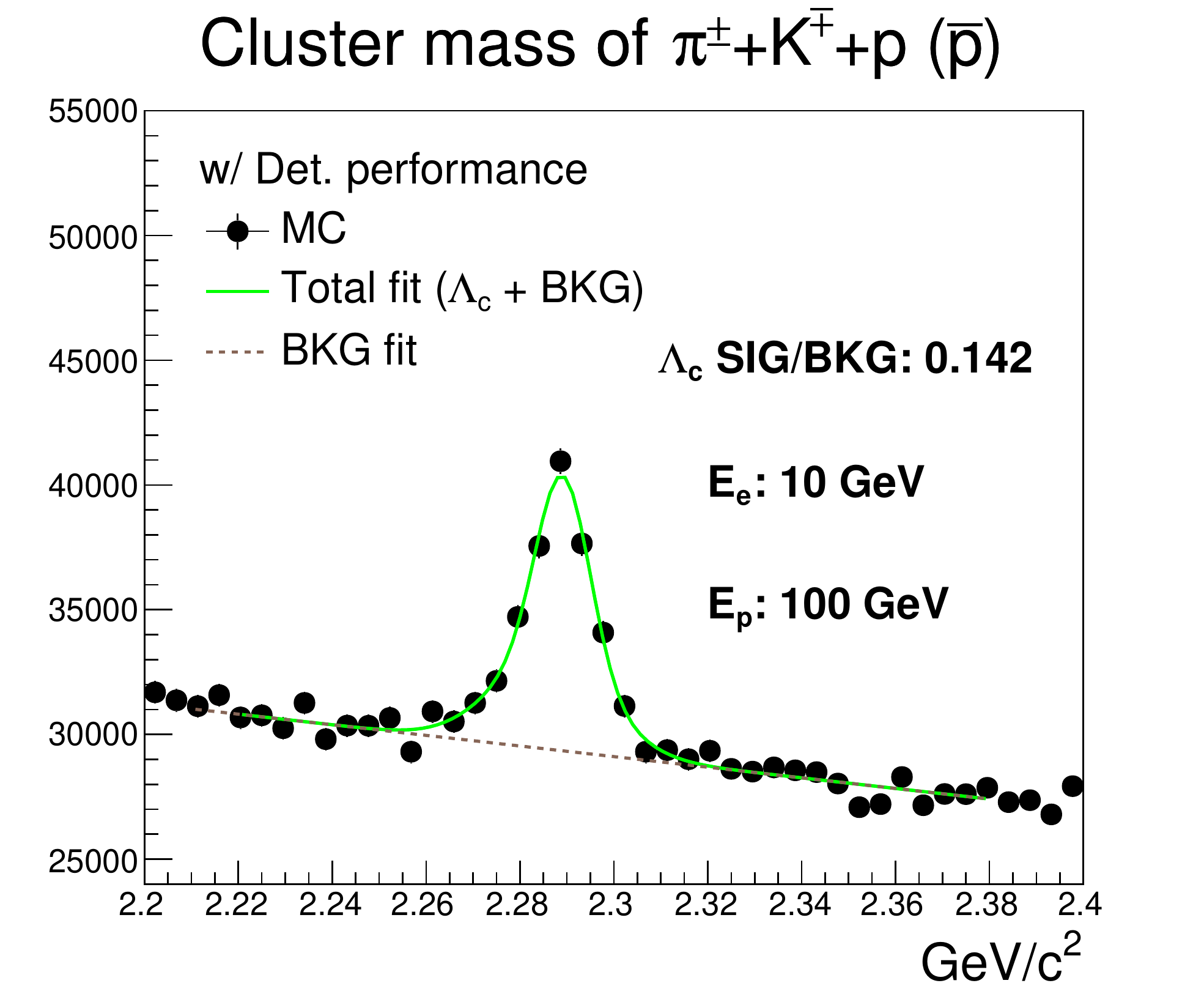}
\includegraphics[width=0.6\textwidth]{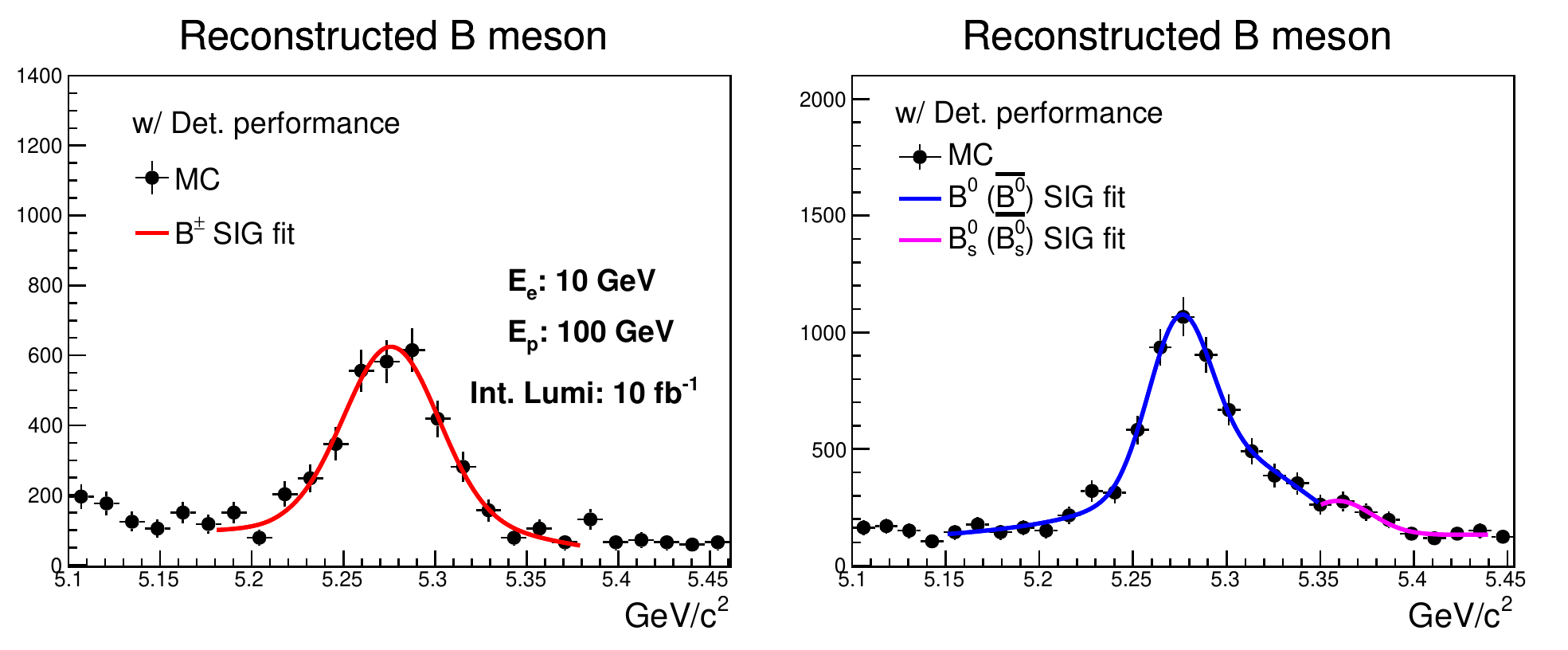}
\caption{Reconstructed heavy flavor hadron mass spectrum with the evaluated silicon tracking detector performance using the Babar magnet (shown in the top row of Figure~\ref{fig:detector_perfom}) in 63.2 GeV $e+p$ collisions with integrated luminosity of 10 $fb^{-1}$.}
\label{fig:rec_HF_mass}
\end{figure}

\begin{figure}[htb!]
\centering
\includegraphics[width=0.6\textwidth]{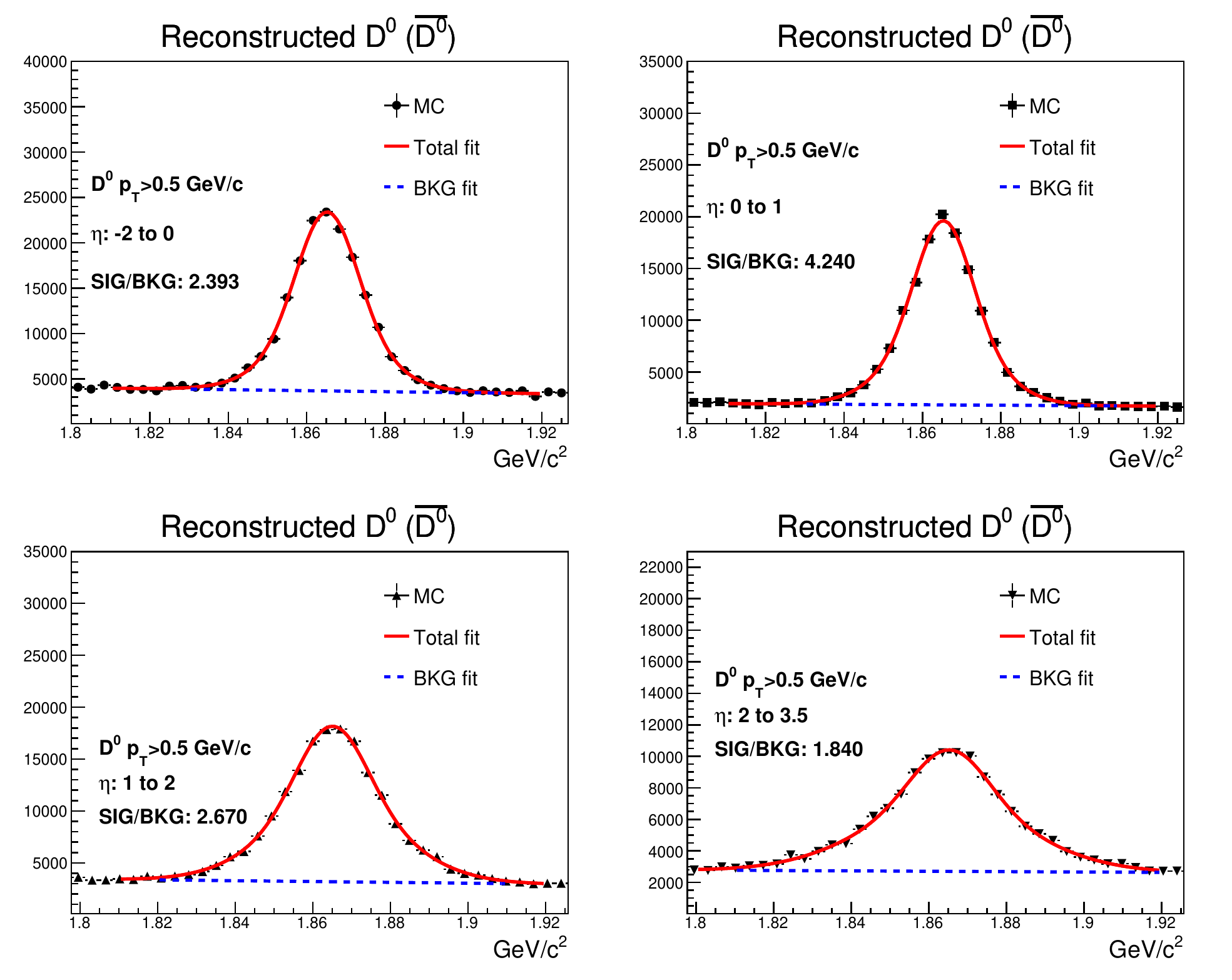}
\caption{Reconstructed $D^{0} (\bar{D^{0}})$ mass spectrum with the evaluated silicon tracking detector performance using the Babar magnet (shown in the top row of Figure~\ref{fig:detector_perfom}) in 63.2 GeV $e+p$ collisions with integrated luminosity of 10 $fb^{-1}$.}
\label{fig:rec_D0_mass}
\end{figure}

For the heavy flavor hadron reconstruction, a series of topological cuts, which include the charged track transverse displaced vertex matching and track transverse momentum threshold constrained by the proposed magnet, have been applied in simulation. The invariant mass distributions of reconstructed heavy flavor hadrons with the detector performance using the Babar magnet in 10 $fb^{-1}$ electron+proton ($e+p$) collisions at 63.2 GeV center of mass energy are illustrated in Figure~\ref{fig:rec_HF_mass}. Good signal over background ratios have been obtained for reconstructed $D^{\pm}$, $D^{0} (\bar{D^{0}})$, $D_{s}^{\pm}$, $\Lambda_{c}^{\pm}$, $B^{\pm}$, $B^{0} (\bar{B^{0}})$ and $B_{s}^{0} (\bar{B_{s}^{0}})$. A better mass resolution and slightly higher signal over background ratios in heavy flavor hadron reconstruction can be obtained with the Beast magnet \cite{lanl_fst}. Figure~\ref{fig:rec_D0_mass} shows the reconstructed $D^{0} (\bar{D^{0}})$ within four different pseudorapidity regions from -2 to 3.5 in 10 $fb^{-1}$ 10 GeV electron and 100 GeV proton collisions with the same simulation configuration applied in Figure~\ref{fig:rec_HF_mass}.

\begin{figure}[htb!]
\centering
\includegraphics[width=0.4\textwidth]{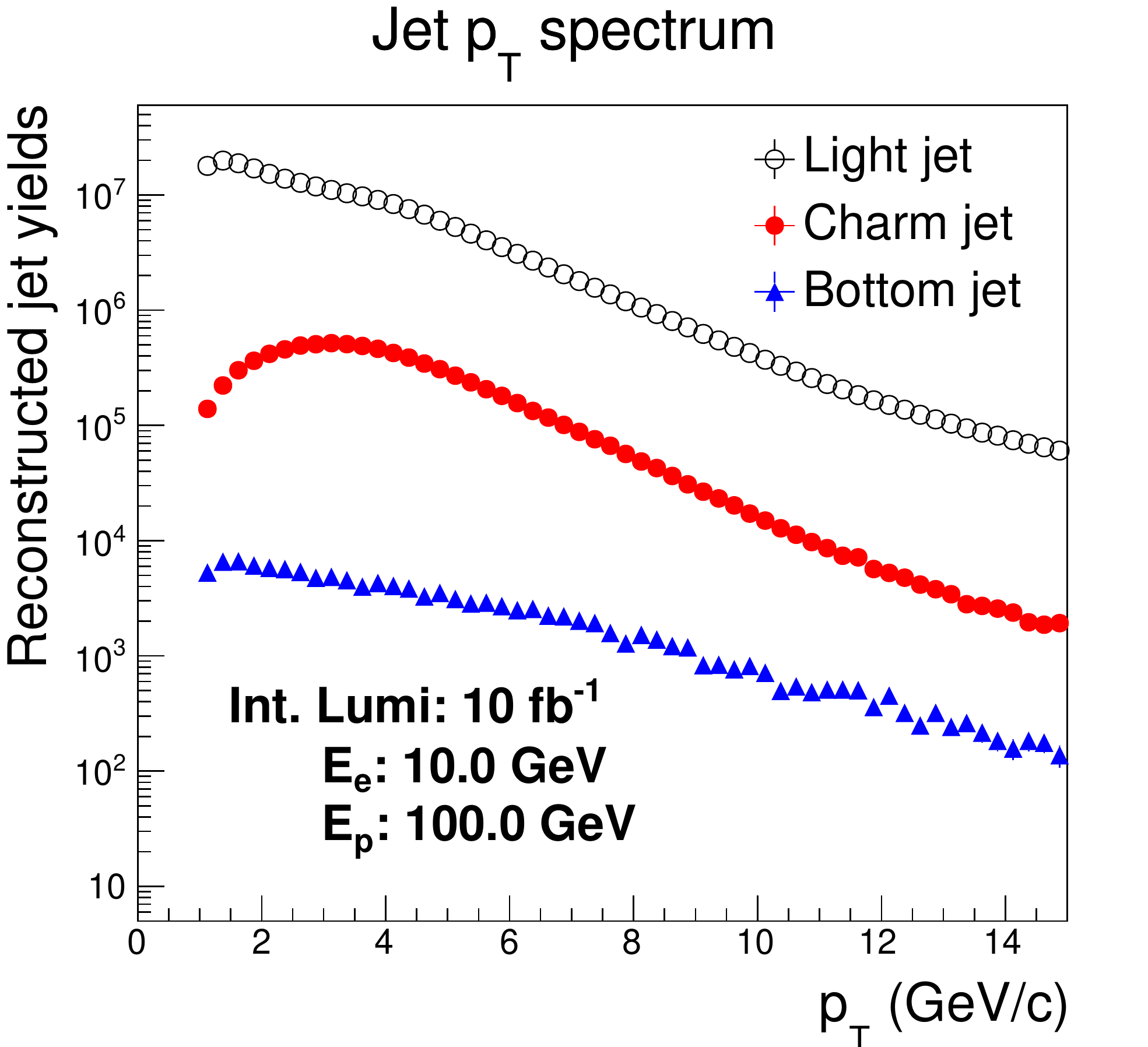}
\caption{Reconstructed jet transverse momentum ($p_{T}$) distributions in 10 GeV electron and 100 GeV proton collisions with integrated luminosity of 10 $fb^{-1}$. Light flavor jets $p_{T}$ spectrum is shown in open black circles, charm jet $p_{T}$ distribution is shown in red solid circles and the bottom jet $p_{T}$ distribution is shown in blue solid triangles.}
\label{fig:rec_jet_pT}
\end{figure}

Jets are reconstructed via the anti-$k_{T}$ algorithm \cite{Cacciari:2008gp} with the cone radius ($R=\sqrt{{\Delta \eta}^{2}+{\Delta \varphi}^{2}}$) set at 1.0 in $e+p$ simulation. The detector response defined in \cite{EIC_YR} has been included in the simulation. Jet flavor is tagged when a fully reconstructed light/heavy flavor hadron is within the jet cone or a jet contains multiple tracks from a charm/bottom hadron decay. The transverse momentum ($p_{T}$) distributions of reconstructed jets with different flavors in 10 $fb^{-1}$ $e+p$ collisions at 63.2 GeV are illustrated in Figure~\ref{fig:rec_jet_pT}. 

\section{Nuclear Modification Factor $R_{eA}$ Projection}
\label{sec:phy_project}
In order to explore the sensitivity to the hadronization in nuclear medium through the proposed hadron and jet measurements at the future EIC, we have carried out simulation studies of flavor dependent nuclear modification factor $R_{eA}$, which is defined in Equation \ref{eq2}, in 10 GeV electron and 100 GeV proton/nucleus collisions.

\begin{equation}
R_{eA} = \frac{1}{A}\frac{d\sigma_{e+A}/dp_{T}d\eta}{d\sigma_{e+p}/dp_{T}d\eta}.
\label{eq2}
\end{equation}

\begin{figure}[htb!]
\centering
\includegraphics[width=0.42\textwidth]{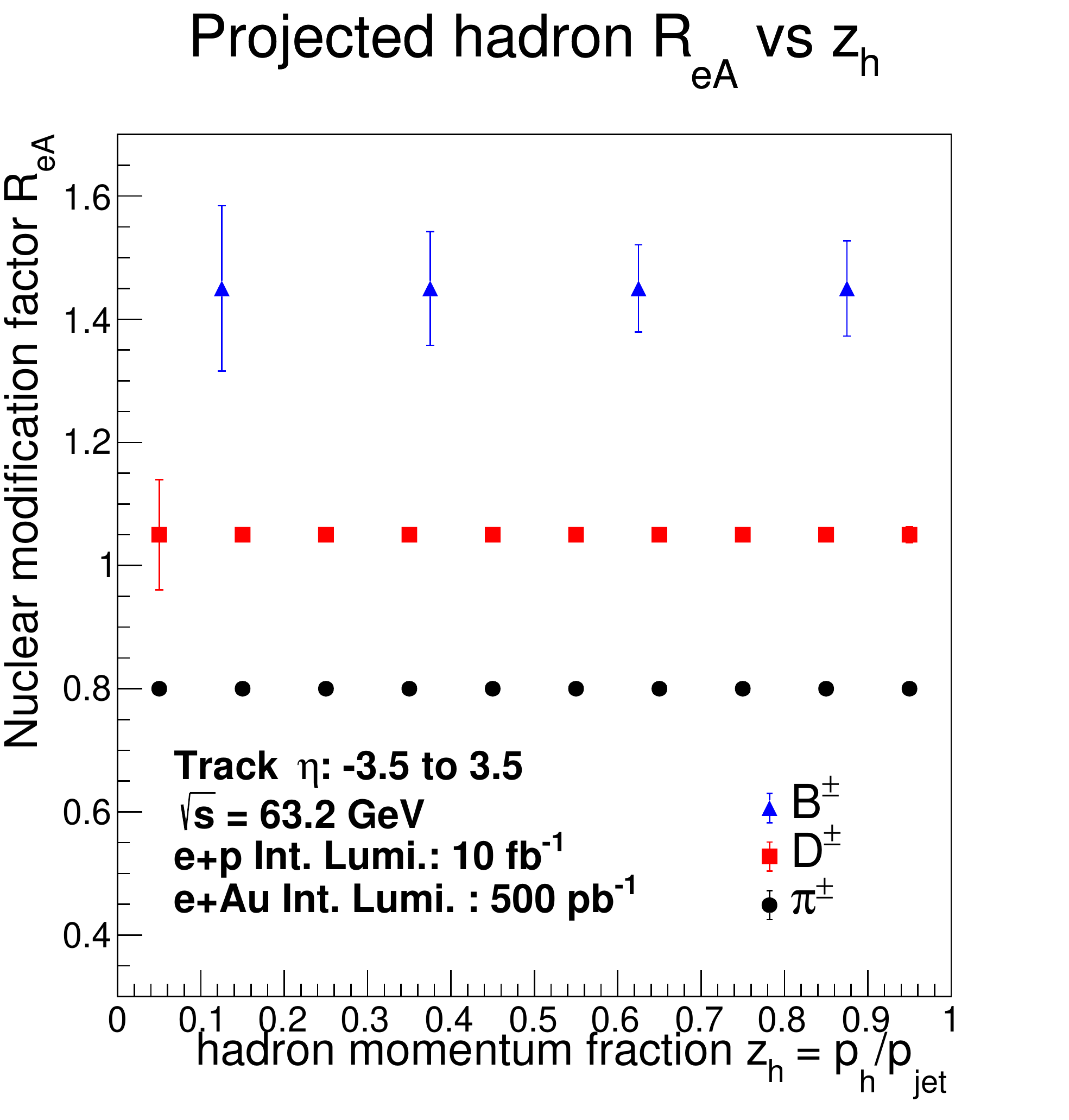}
\includegraphics[width=0.42\textwidth]{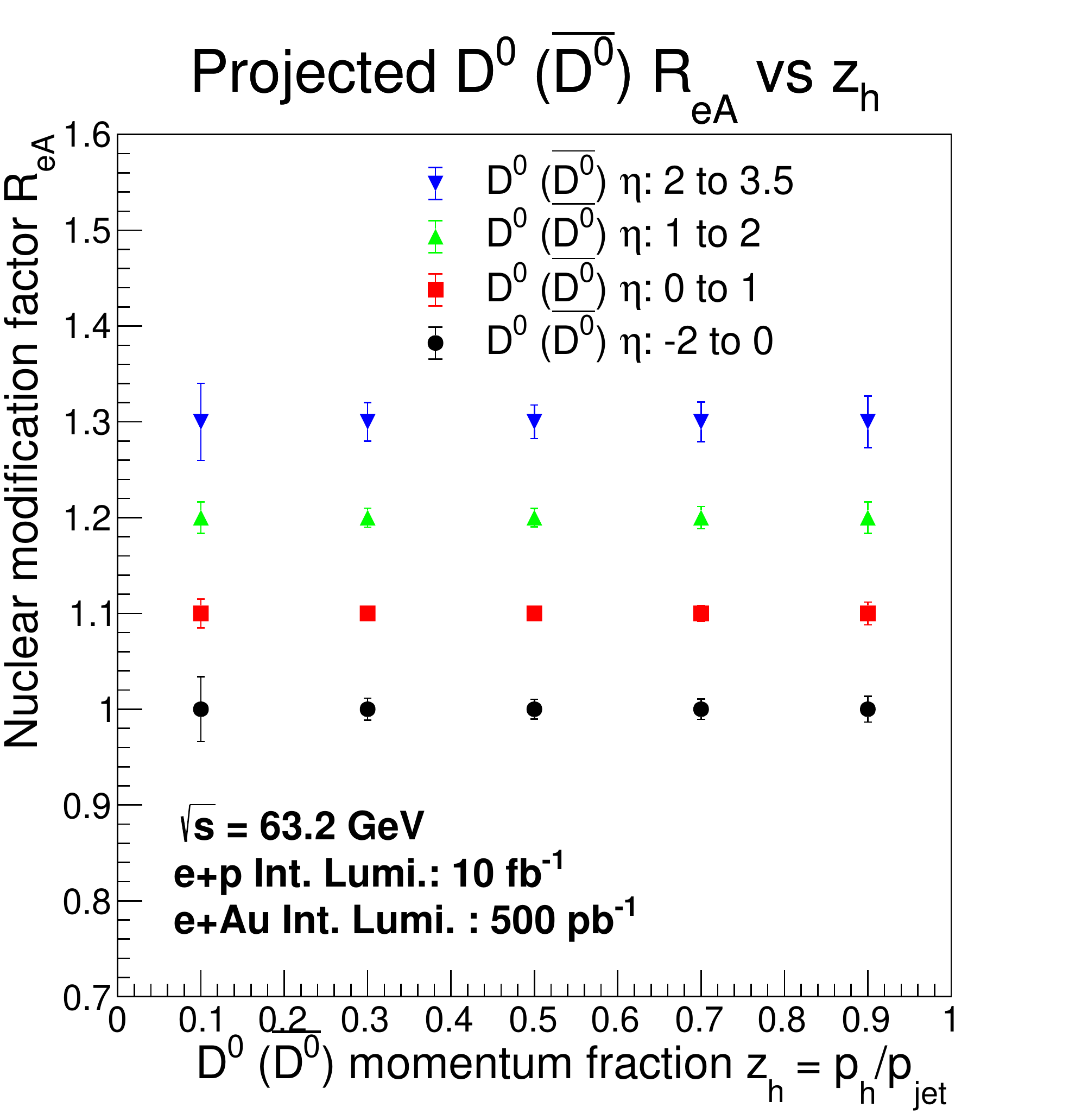}
\caption{Flavor dependent nuclear modification factor $R_{eAu}$ for reconstructed hadrons with different flavors (left) and pseudorapidity separated reconstructed $D^{0} (\bar{D^{0}})$ $R_{eAu}$ (right) in 10 $fb^{-1}$ $e+p$ collisions and 500 $pb^{-1}$ $e+Au$ collisions at $\sqrt{s} = 63.2$ GeV. The reconstructed hadron yields are extracted from PYTHIA simulation with the proposed silicon vertex/tracking detector performance within the Babar magnet.}
\label{fig:flavor_ReA}
\end{figure}

Hadron yields in $e+p$ collisions, $d\sigma_{e+p}/dp_{T}d\eta$, are extracted from reconstructed heavy/light flavor hadron mass spectrums discussed in Section \ref{sec:reco}. The yields in $e+Au$ collisions, $d\sigma_{e+Au}/dp_{T}d\eta$, are scaled from the corresponding cross-section in $e+p$ collisions at the same collision energy by the nuclear mass number $A$. The left panel of Figure~\ref{fig:flavor_ReA} illustrates the statistical projection of hadron momentum fraction $z_{h}$, which is the ratio of reconstructed hadron momentum over associated jet momentum, dependent nuclear modification factor $R_{eAu}$ for reconstructed $\pi^{\pm}$, $D^{\pm}$ and $B^{\pm}$ in 10 $fb^{-1}$ $e+p$ collisions and 500 $pb^{-1}$ $e+Au$ collisions at 63.2 GeV center of mass energy. 

The projection of hadron momentum fraction dependent $R_{eAu}$ for reconstructed $D^{0} (\bar{D^{0}})$ in pseudorapidity regions of -2 to 0, 0 to 1, 1 to 2 and 2 to 3.5 using 10 $fb^{-1}$ $e+p$ collisions and 500 $pb^{-1}$ $e+Au$ collisions at $\sqrt{s} = 63.2$ GeV is shown in the right panel of Figure~\ref{fig:flavor_ReA}. Less than 10$\%$ statistical uncertainties can be obtained by reconstructed heavy flavor $R_{eAu}$ measurements with around one year EIC operation. These proposed measurements will provide better understanding in the probed nuclear PDFs and fragmentation functions in nuclear medium in a wide kinematic region.

\section{Conclusion}
The future EIC will create an ideal QCD environment to explore the hadronization process in vacuum and the nuclear medium through a series of heavy flavor and jet measurements. Clear heavy flavor signals have been obtained in simulation studies with the performance of the proposed EIC detector concept, which includes a silicon vertex/tracking subsystem. Good precision of the flavor dependent nuclear modification factor projection at the EIC will provide a strong discriminating power to separate different model predictions such as the parton energy loss mechanism \cite{Li:2020zbk}.

\section*{Acknowledgements}


\paragraph{Funding information}
This work is supported by the Los Alamos National Laboratory LDRD project 20200022DR project.

\bibliography{DIS2021_XuanLi.bib}

\begin{thebibliography}{1}
\providecommand{\url}[1]{\texttt{#1}}
\providecommand{\urlprefix}{URL }
\expandafter\ifx\csname urlstyle\endcsname\relax
  \providecommand{\doi}[1]{doi:\discretionary{}{}{}#1}\else
  \providecommand{\doi}{doi:\discretionary{}{}{}\begingroup
  \urlstyle{rm}\Url}\fi
\providecommand{\eprint}[2][]{\url{#2}}

\bibitem{HF_had_th}
E.~Norrbin and T.~Sjostrand,
\newblock \emph{{Production and hadronization of heavy quarks}},
\newblock Eur. Phys. J. C \textbf{17}, 137 (2000),
\newblock \doi{10.1007/s100520000460},
\newblock \eprint{hep-ph/0005110}.

\bibitem{EIC_YR}
R.~Abdul~Khalek \emph{et~al.},
\newblock \emph{{Science Requirements and Detector Concepts for the
  Electron-Ion Collider: EIC Yellow Report}} (2021),
  \eprint{http://arxiv.org/abs/2103.05419}.

\bibitem{lanl_fst}
C.-P. Wong, X.~Li, M.~Brooks, M.~J. Durham, M.~X. Liu, A.~Morreale, C.~da~Silva
  and W.~E. Sondheim,
\newblock \emph{{A Proposed Forward Silicon Tracker for the Future Electron-Ion
  Collider and Associated Physics Studies}} (2020),
  \eprint{https://arxiv.org/abs/2009.02888}.

\bibitem{Arrington:2021yeb}
J.~Arrington \emph{et~al.},
\newblock \emph{{EIC Physics from An All-Silicon Tracking Detector}}  (2021),
\newblock \eprint{2102.08337}.

\bibitem{Sjostrand:2007gs}
T.~Sjostrand, S.~Mrenna and P.~Z. Skands,
\newblock \emph{{A Brief Introduction to PYTHIA 8.1}},
\newblock Comput. Phys. Commun. \textbf{178}, 852 (2008),
\newblock \doi{10.1016/j.cpc.2008.01.036},
\newblock \eprint{0710.3820}.

\bibitem{AGOSTINELLI2003250}
S.~Agostinelli \emph{et~al.},
\newblock \emph{Geant4—a simulation toolkit},
\newblock Nuclear Instruments and Methods in Physics Research Section A:
  Accelerators, Spectrometers, Detectors and Associated Equipment
  \textbf{506}(3), 250 (2003),
\newblock \doi{https://doi.org/10.1016/S0168-9002(03)01368-8}.

\bibitem{Cacciari:2008gp}
M.~Cacciari, G.~P. Salam and G.~Soyez,
\newblock \emph{{The anti-$k_t$ jet clustering algorithm}},
\newblock JHEP \textbf{04}, 063 (2008),
\newblock \doi{10.1088/1126-6708/2008/04/063},
\newblock \eprint{0802.1189}.

\bibitem{Li:2020zbk}
H.~T. Li, Z.~L. Liu and I.~Vitev,
\newblock \emph{{Heavy meson tomography of cold nuclear matter at the
  electron-ion collider}},
\newblock Phys. Lett. B \textbf{816}, 136261 (2021),
\newblock \doi{10.1016/j.physletb.2021.136261},
\newblock \eprint{2007.10994}.

\end{thebibliography}

\nolinenumbers

\end{document}